\documentclass{article}
\usepackage{slashed,epsfig,amsmath,amssymb,enumitem}
\usepackage[papersize={8.5in,11in}]{geometry}
   \usepackage{bm}
\usepackage{graphicx}
\newcommand{\be}{\begin{equation}}
\newcommand{\ee}{\end{equation}}

\begin{document}
\title{Electroweak Symmetry Breaking}
\author{J. W. Moffat\\
Perimeter Institute for Theoretical Physics, Waterloo, Ontario N2L 2Y5, Canada
and\\
Department of Physics and Astronomy, University of Waterloo, Waterloo,\\
Ontario N2L 3G1, Canada}
\maketitle


\begin{abstract}
This paper examines the Higgs particle self-coupling and its implications for electroweak symmetry breaking in the Standard Model. We review the current experimental constraints on the Higgs trilinear coupling and discuss the challenges in measuring it precisely. The potential consequences of deviations from the Standard Model prediction are explored, including the possibility of new physics. We then consider an alternative ultraviolet complete electroweak theory based on finite quantum field theory, which does not require spontaneous symmetry breaking or a non-zero vacuum expectation value. The predictions of this model for particle masses and the stability of the electroweak vacuum are compared to the standard Higgs mechanism. Finally, we review the SM derivation of the W-boson mass, its dependence on radiative corrections and its experimentally determined value.
\end{abstract}

\section{Introduction}

The discovery of the Higgs boson in 2012 marked a major milestone in particle physics, confirming a key prediction of the Standard Model (SM) and providing insight into the mechanism of electroweak symmetry breaking. However, important questions remain about the nature of the Higgs field and its self-couplings. The shape of the Higgs potential, determined by these self-couplings, is crucial for understanding electroweak symmetry breaking and the stability of the vacuum. In the SM, the Higgs trilinear coupling is precisely predicted based on the Higgs mass and vacuum expectation value. Measuring this coupling experimentally would provide a critical test of the Higgs mechanism. However, direct observation of Higgs self-interactions is extremely challenging due to the small amplitudes involved. Current experimental bounds from the LHC remain inconclusive.

This paper examines the theoretical and experimental status of Higgs self-couplings and their implications. We begin by reviewing the SM predictions and current experimental constraints. The difficulties in measuring the trilinear self-coupling of the Higgs particle precisely at present and future colliders are discussed. We then explore the potential consequences if deviations from the SM value are observed, including possible indications of physics beyond the SM. Given the lack of direct evidence for Higgs self-interactions, we consider an alternative approach to electroweak symmetry breaking based on finite quantum field theory. This ultraviolet complete theory does not require a non-zero vacuum expectation value or spontaneous symmetry breaking. We compare its predictions for particle masses and vacuum stability to those of the conventional Higgs mechanism.

We review the SM derivation of the W-boson mass, examining its dependence on radiative corrections and experimental determinations of its mass. This provides context for comparing the predictions of different electroweak models. By exploring these topics, we aim to elucidate the current state of understanding of electroweak symmetry breaking and highlight key areas for future theoretical and experimental investigation.

\section{Symmetry breaking and Higgs particle self-coupling}

The SM electroweak sector has been formulated as a spontaneous breaking of the gauge invariant symmetry group $SU(2)\times U(1)$ ~\cite{Higgs1,Englert,Guralnik,Higgs2,Kibble,Halzen,Burgess}.  
The Lagrangian for the Higgs field takes the form:
\be
\mathcal{L}_{\text{Higgs}} = (D_\mu \phi)^\dagger (D^\mu \phi) - V(\phi),
\ee
where $D_\mu$ is the covariant derivative:
\be
D_\mu = \partial_\mu + ig\frac{\tau^a}{2}W_\mu^a + ig'\frac{Y}{2}B_\mu.
\ee
The $W_\mu^a$ fields where a=1,2,3 are the gauge fields associated with the $SU(2)_L$ weak isospin gauge group in the SM. These are vector fields that mediate the weak interaction and  $W_\mu^{\pm} = \frac{1}{\sqrt{2}}(W_\mu^1 \mp iW_\mu^2)$. 
The $B_\mu$ field is the gauge field associated with the $U(1)_Y$ weak hypercharge gauge group in the SM. The $\phi$ is the complex SU(2) doublet Higgs field 
\be
\phi = \begin{pmatrix} \phi^+ \\ \phi^0 \end{pmatrix}.
\ee
The Higgs mechanism, proposed to explain the origin of particle masses, is a cornerstone of the SM of particle physics. Central to this mechanism is the Higgs potential, given by
\be
V(\phi)=\mu^2\phi^\dagger\phi 
+\lambda(\phi^\dagger\phi)^2,
\ee
where $\mu^2<0$ for spontaneous symmetry breaking. The Higgs field acquires after spontaneous symmetry breaking a non-zero vacuum expectation value:
\be
\langle 0|\phi| 0\rangle = \frac{1}{\sqrt{2}}\begin{pmatrix} 0 \\ v \end{pmatrix},
\ee
where $v = \sqrt{-\mu^2/\lambda}$ is the vacuum expectation value.
After electroweak symmetry breaking, $W^3_\mu$ and $B_\mu$ mix to form the physical Z boson and the photon field $A_\mu$:
\be
Z_\mu = \cos\theta_w W_\mu^3 - \sin\theta_w B_\mu,\quad  A_\mu = \sin\theta_w W_\mu^3 + \cos\theta_w B_\mu,
\ee
where $\theta_w$ is the weak mixing angle.

After spontaneous symmetry breaking of $SU(2)\times U(1)$, we can express the Higgs field in terms of the physical Higgs boson and the Goldstone bosons that are eaten by the gauge bosons. In the unitary gauge, the Higgs doublet becomes:
\be
\phi = \frac{1}{\sqrt{2}} \begin{pmatrix} 0 \\ v + \chi(x) \end{pmatrix}.
\ee
where $\chi$ is the neutral scalar Higgs field. The Higgs potential potential takes the form:
\be
\label{symmetrybreaking}
V(\chi)=\frac{1}{2}\mu^2\chi^2+\lambda_3 v \chi^3+\lambda \chi^4.
\ee

To experimentally verify the spontaneously broken Higgs potential, we need to measure the Higgs self-interaction. The most promising approach is through the measurement of the trilinear Higgs coupling $\lambda_3$. However, direct processes like $H\rightarrow 3H$ or $H\rightarrow 4H$ are kinematically forbidden due to energy conservation. The most feasible method to probe the Higgs self-coupling at current colliders is through loop-induced processes. The primary channel is gluon fusion: $gg\rightarrow H^*\rightarrow  HH$ described by Fig. 1. This process involves two gluons connected to a top quark triangle loop, producing a virtual Higgs that decays into two on-shell Higgs bosons. The amplitude for this process is approximately 1200 times smaller than the single Higgs production cross-section, making it extremely challenging to measure at the LHC. The quadilinear Higgs self-coupling $\lambda \chi^4$ cannot be measure at present accelerator energies. Even at an energy $\sqrt{s}=100$ TeV the background interference would make the detection difficult to determine.

The best measurement from the CMS experiment at CERN constrains at $\sqrt{s}=13$ TeV the ratio $\kappa_{\lambda}=\lambda_3/\lambda_{3SM}$ to $ - 1.2 < \kappa_\lambda < +7.5$, where $\kappa_\lambda=1$ corresponds to the SM prediction~\cite{CMS}.

In an electroweak symmetry breaking model $\lambda_3$ can be a free parameter not necessarily tied to the Higgs particle mass. It could be enhanced by a factor $\beta$ compared to the SM value, $\lambda_3 =\beta\lambda_{3SM}$. Measuring $\lambda_3$ through e.g., di-Higgs particle production and comparing it to the SM prediction can test the nature of electroweak symmetry breaking. A significant deviation from the SM value would indicate either physics beyond the SM, such as new particles or a different dynamical symmetry breaking scenario.

Present measurements are consistent with $\kappa_\lambda\sim 0$, which implies no direct experimental confirmation of the Higgs mechanism as the source of particle masses. This lack of evidence poses a significant challenge to our understanding of electroweak symmetry breaking. To measure $\kappa_\lambda$ with sufficient precision would require a proton-proton collider operating at energies around 100 TeV. Such a facility, often referred to as a Future Circular Collider (FCC), is under consideration but not yet approved or constructed. The inability to measure the Higgs self-coupling precisely leaves open questions about the nature of electroweak symmetry breaking. Alternative theories, such as composite Higgs models or extended Higgs sectors, cannot be ruled out based on current data. While the SM predicts a specific value for the Higgs self-coupling, beyond-the-Standard-Model theories predict deviations. Precise measurements of $\kappa_\lambda$ could provide crucial insights into new physics.

\begin{figure}
    \centering
    \includegraphics[width=0.75\linewidth]{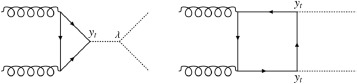}
    \caption{Higgs particle self-coupling di-pair loop diagrams}
    \label{fig:enter-label}
\end{figure}

The experimental determination of the Higgs particle self-coupling will decide whether the standard electroweak model with spontaneous breaking of $SU(2)\times U(1)$ is the correct model, or whether an alternative model is required to replace it. 

\section{Ultraviolet complete electroweak theory}

If the self-coupling of the Higgs particle is found not to agree with the SM shape of the Higgs field potential, then we can consider replacing the standard renormalizable theory with an alternative finite formulation of the electroweak sector. The finite quantum field theory 
(QFT)~\cite{Moffat1,Moffat2,Moffat3,Moffat4,Moffat5,Moffat6,Green,Moffat7,Moffat8,Buoninfante,Nortier,
Modesto1,Modesto2,Koshelev,Chin,Tomboulis}, realizes a Poincar\'e invariant, unitary QFT and all quantum loop graphs are ultraviolet finite to all orders of perturbations theory. Although the field operators and the interactions of particles are nonlocal, the model satisfies microscopic causality~\cite{Moffat5,Moffat6}. 
The finite QFT model allows for an alternative interpretation of the electroweak $SU(2)\times U(1)$ sector. Because the loop graphs are finite an infinite renormalization of particle interactions is not required. This allows for finite field theory interactions with massive bosons and fermions in the Lagrangian. The idea that the $SU(2)\times U(1)$ Lagrangian has initially to be massless at the outset to guarantee a gauge invariant and renormalizable scheme is not adopted as a basis for the alternative model.

The boson and fermion masses are calculated from perturbative one-loop graphs with an associated QFT length (mass) scale $\Lambda_i$. By determining the scale $\Lambda_{WZ}$ for the Z-boson mass, the mass of the W boson is predicted to be $M_W=80.05$ GeV close to the observed mass $M_W=80.379\pm 0.012$ GeV~\cite{Moffat6}.  The fermion particle masses are also calculated from perturbative loop graphs with associated values of $\Lambda_i$~\cite{Moffat6}. The Higgs particle mass $M=125$ GeV is calculated and the Higgs mass hierarchy fine-tuning problem is resolved. All the low energy predicted decay products and particle productions verified by the LHC will be retained in the alternative model for energy scales less than $\Lambda_i$ of order TeV. However, the only electroweak true vacuum will be $v=0$ predicting a stable vacuum in contrast to the standard prediction by the Higgs mechanism of an unstable vacuum at very high energies~\cite{Kusenko}.

The SM spontaneous symmetry breaking mechanism with the non-vanishing vev, $v=\langle 0|\phi|0\rangle$, predicts the triple and quartic self-interaction terms in the potential $V(\phi)$ after symmetry breaking Eq. (\ref{symmetrybreaking}).
In the SM, the assumed specific self-coupling term after symmetry breaking, $\lambda_3 v\chi^3$, predicts the relation between the self-coupling constant $\lambda$ and the vacuum expectation value $v$, $M_H=\sqrt{2\lambda}v$. The vacuum expectation value is known from the Fermi theory, 
\be
v=1/(\sqrt{2}G_F)^{1/2}\sim 246.3 {\rm GeV},
\ee
leading to the predicted value $\lambda\sim 0.13$. A measured modification of the Higgs potential will modify these predictions. Only a direct measurement of the Higgs field self-coupling interaction can determine the shape of the Higgs field potential.

\section{Electroweak low energy predictions}

For the energy scale $M_W << \Lambda_{\rm EW}$, where $\Lambda_{\rm EW}$ is a high-energy electroweak energy scale, low energy electroweak predictions can be derived. An important experimental prediction is the derivation of the W-boson mass. The W-boson mass derivation from the standard spontaneously broken $SU(2)\times U(1)$ model is given by
\be
M_W = \frac{1}{2}gv.
\ee
The predicted Z-boson mass is
\be
M_Z = \frac{1}{2}\sqrt{g^2+{g'}^2},
\ee
where g is the coupling constant associated with the $SU(2)_L$ weak isospin gauge group
and $g^\prime$ is the coupling constant associated with the $U(1)_Y$ weak hypercharge gauge group. We have $e = g\sin\theta_w = g'\cos\theta_w$ where e is the electric charge.

Instead of following the SM derivation of the W-boson mass, from the spontaneously broken $SU(2)\times U(1)$ model, the W mass can be derived from low energy effective field theory~\cite{Sirlin,Awramik}. The local four-fermion Fermi V-A effective theory is matched to the electroweak theory in the low energy limit of muon decay. The Lagrangian is given by
\be
{\cal L}_{V-A}= -\frac{G_F}{\sqrt{2}}{\bar \psi}_{\nu_\mu}\gamma^\mu(1-\gamma_5)
\psi_\mu {\bar \psi}_e\gamma_\mu(1-\gamma_5)\psi_{\nu_e},
\ee
where the Fermi constant is determined by
\be
\frac{G_F}{\sqrt{2}}= \frac{{\hat g}^2}{8M_W^2},
\ee
and 
\be
{\hat g}^2=g^2(1+\Delta r).
\ee
The $\Delta r$ denotes radiative corrections.

The constant $G_F$ is related to the $\mu$ particle lifetime by the formula:
\be
\frac{1}{\tau}= \frac{G_F^2 m_\mu^5}{192\pi^3}\biggl(1-\frac{8m_e^2}{m_\mu^2}\biggr)
\biggl[1+\frac{3}{5}\frac{m_\mu^2}{M_W^2}+\frac{\alpha}{2\pi}\biggl(\frac{25}{4}-\pi^2\biggr)\biggr].
\ee

A prediction of the W-boson mass is given by ~\cite{Sirlin}:
\be
\label{Wmass}
M_W=\biggl(\frac{\pi\alpha}{\sqrt{2}G_F}\biggr)^{1/2}\frac{(1+\Delta r)}{\sin\theta_w}.
\ee
The one-loop radiative correction can be written:
\be
\Delta r= \Delta\alpha - \frac{c_W^2}{s_W^2}\Delta\rho +\Delta r_{\rm rem}(M_H),
\ee
where $c_W^2=\cos^2\theta_w=M_W^2/M_Z^2$ and $s_W^2=1-c_W^2$. Moreover, $\Delta \rho$ is the radiative corrected parameter $\rho$:
\be
\rho= \frac{M_W^2}{c_W^2M_Z^2}.
\ee

The prediction of the $W$ boson in Eq. (\ref{Wmass}) does not depend at the classical tree graph level on the field theory assumption about the Higgs scalar neutral particle. However, the calculation of the radiative corrections $\Delta r$ depends logarithmically on the Higgs mass, $M_H\sim 125$ GeV, and contributes negligibly to $\Delta r$.

Fermilab's most recent measurement of the W-boson mass ~\cite{Abulencia} was conducted using data from the Collider Detector at Fermilab (CDF) and resulted in a mass value of $80.4335\pm 0.0094$ GeV/c2. This value is higher than the SM prediction and previous experimental results. The CERN ATLAS collaboration has also measured the W-boson mass~\cite{ATLAS}. The reported W-boson mass is: $80.370\pm 0.019$ GeV/c2. This value is closer to the SM prediction.

\section{Conclusions}

The experimental verification of the Higgs self-coupling remains one of the most pressing challenges in particle physics. It is crucial for confirming our understanding of electroweak symmetry breaking and the origin of particle masses. Future high-energy colliders and advanced detection techniques will be essential for progress in this area. This overview highlights the current state of experimental efforts to verify the Higgs self-coupling and the challenges faced in probing the fundamental nature of electroweak symmetry breaking. It underscores the need for continued research and development in high-energy physics to address these fundamental questions about the nature of our universe.

\section*{Acknowledgments}

Thanks are given to Viktor Toth and Martin Green for helpful discussions. This research was supported in part by Perimeter Institute for Theoretical Physics. Research at Perimeter Institute is supported by the Government of Canada through the Department of Innovation, Science and Economic Development Canada and by the Province of Ontario through the Ministry of Research, Innovation and Science.

\end{document}